\documentclass[twocolumn,aps,showpacs]{revtex4}

\usepackage{color}

\usepackage{graphicx}
\bibpunct{}{}{,}{s}{}{}

\begin{document}
\title{First-principles calculations of strontium on Si(001)} 
\author{Christopher R. Ashman,$^{1}$ Clemens J. F\"orst,$^{1,2}$
Karlheinz Schwarz$^{2}$ and Peter E. Bl\"ochl,$^{1,*}$}
\affiliation{$^1$ Clausthal University of Technology, Institute for
Theoretical Physics, Leibnizstr.10, D-38678 Clausthal-Zellerfeld,
Germany}
\affiliation{$^2$ Vienna University of Technology, Institute for
Materials Chemistry, Getreidemarkt 9/165-TC, A-1060 Vienna, Austria}
\date{\today} 
\begin{abstract}
This paper reports state-of-the-art electronic structure calculations on
the deposition of strontium on the technologically relevant, (001)
orientated silicon surface.  We identified the surface reconstructions
from zero to four thirds monolayers and relate them to experimentally
reported data. A phase diagram is proposed.  We predict phases at 1/6,
1/4, 1/2, 2/3 and 1 monolayers.  Our results are expected to provide
valuable information in order to understand heteroepitaxial growth of a
prominent class of high-K oxides around SrTiO$_3$.  The insight obtained
for strontium is expected to be transferable to other alkaline earth
metals.  
\end{abstract}
\maketitle

\section{Introduction}

Device scaling has been the engine driving the microelectronics
revolution as predicted by Moore's law.\cite{Moore95} By reducing the
size of transistors, processors become faster and more power efficient
at an exponential rate. Currently the main challenge in device
scaling is the integration of high-K oxides as gate oxides into
silicon technology. 

The gate oxide is the dielectric of a capacitor, which is used to
attract charge carriers into the channel region.  Thus a current can
flow from source to drain, provided a voltage is applied to the gate
electrode.  With a thickness of only 1-2 nm\cite{roadmap} the gate
dielectric is the smallest structure of a transistor.  As the thickness
of the gate oxide is further reduced, its insulating property is lost
due to direct tunneling through the ultrathin oxide. The results are
intolerable leakage currents and a large power consumption.

A remedy to this problem is the replacement of the current SiO$_2$ based
gate oxides with an insulator having a larger dielectric constant, a
so-called high-K oxide.  A high-K oxide gate with the same capacity as
an ultrathin SiO$_2$ based one  will be thicker and should therefore
exhibit smaller leakage currents due to direct tunneling.  The
integration of new oxides into the semiconductor technology has,
however, proven to be a major problem.  Hence an enormous research
effort is underway to understand growth of high-K oxides onto silicon.

Currently, HfO$_2$ and ZrO$_2$ are the main contenders for the first
generation of high-K oxides to be introduced in fabrication. These
oxides still exhibit an interfacial SiO$_2$ layer and therefore do not
form a direct interface with silicon. As scaling proceeds, an
interfacial SiO$_2$ layer cannot be tolerated anymore.  The existence of
an atomically abrupt interface between silicon and a high-K oxide has
been demonstrated by McKee et al.\cite{McKee98,McKeeScience} for
Ba$_x$Sr$_{1-x}$TiO$_3$ on Si(001), after an epitaxial relationship has
been reported in the late 80's.\cite{Ishiwara88,Mori91}

A detailed understanding of metal adsorption is crucial to control oxide
growth on Si. The growth process is guided by the sequence of structures
that develop as the metal is deposited on the silicon
surface.\cite{McKee98} The nature of these structures as well as the
interface between Si and a high-K oxide is, however, still under
debate.

The adsorption of the alkaline earth metals Sr and Ba on Si(001) has
been extensively studied. Most of the studies of Sr on Si(001) are
diffraction studies such as low-energy electron diffraction
(LEED)~\cite{Fan90,Bakhtizin96-1,Bakhtizin96-2,Hu01,Herrera01,Liang01}
and reflection high-energy electron diffraction
(RHEED)~\cite{McKee93,Lettieri02,Norton02,McKeeScience03} or scanning
tunneling microscopy (STM)~\cite{Bakhtizin96-1,Bakhtizin96-2,Liang01}
experiments. The STM studies have been most valuable because they
contribute detailed real-space information on the atomic scale.  Similar
LEED,\cite{Urano96,Takeda98,Hu99} RHEED~\cite{McKee91} and
STM~\cite{Yao99,Hu00,Ojima01,Ojima02,Ojima02-1} as well as x-ray
photoemission studies~\cite{Cheng98,Hu99} have been performed for Ba.
X-ray standing wave experiments provide valuable restrictions on the
structures with coverages of 1/2 and 1/3 monolayer (ML).\cite{Herrera00}
The photoemmision (XPS)~\cite{Mesarwi90,Herrera01} studies provide
insight into the ionization state of Sr, and show a qualitative change
of the Fermi-level pinning as a function of coverage.\cite{Herrera01}

Diffraction studies suffer from the fact that they average over several
structures and terraces. Here STM experiments provide valuable clues.
One of the major experimental difficulties is the determination of the
coverage at which the data are collected.\cite{Herrera01} 

Theoretical investigations of isolated Ba atoms adsorbed on Si(001) have
been performed by Wang et al.\cite{Wang99}

In this work we address the deposition of Sr on Si(001) using
state-of-the-art electronic structure calculations.  We attempt to
provide a complete set of adsorption structures, their energetics,
chemical binding and electronic structure. We will categorize the
reconstructions by pointing out the driving forces that lead to the various
ordered structures. This provides a unified picture of Sr adsorption
from low coverage up to 4/3 monolayers.

\section{Computational details}
\label{sec:details}

The calculations are based on density functional theory\cite{Kohn,
KohnSham} using a gradient corrected functional.\cite{PBE}  The
electronic structure problem was solved with the projector augmented
wave (PAW) method,\cite{PAW94} an all-electron electronic structure
method using a basis set of plane waves augmented with partial waves
that incorporate the correct nodal structure. The frozen core states
were imported from the isolated atom. For the silicon atoms we used a
set with two projector functions per angular momentum for $s$ and
$p$-character and one projector per angular momentum with $d$-character.
The hydrogen atoms of the back surface had only one $s$-type projector
function.  For strontium we treated the 4$s$ and 4$p$ core shells as
valence electrons.  Per angular momentum we used three $s$-type and two
$p$- and $d$-type projector functions. The augmentation charge density
has been expanded in spherical harmonics up to $\ell=2$. The kinetic
energy cutoff for the plane wave part of the wave functions was set to
30~Ry and that for the electron density to 60~Ry.

A slab of five silicon layers was used as silicon substrate.  Wang et
al.\cite{Wang99} report that the adsorption energy of a Sr atom on the
surface changes by 0.05~eV when between a 4-layer slab and a 6-layer
slab of silicon. In our calculations the energy per additional silicon
atom agrees to within 0.06~eV with that of bulk silicon between a
4-layer and a 5-layer slab. The dangling bonds of the unreconstructed
back surface of the slab have been saturated by hydrogen atoms. The
lateral lattice constant was chosen as the experimental lattice constant
$a=5.4307~$\AA\ of silicon.\cite{CRC} The slabs repeat every 16~\AA\
perpendicular to the surface, which results in a vacuum region of
9.5~\AA\ for the clean silicon surface.

The Car-Parrinello ab-initio molecular dynamics\cite{Car85} scheme with
damped motion was used to optimize the electronic and atomic structures.
All structures were fully relaxed without symmetry constraints.  The
atomic positions of the backplane of the slab and the terminating
hydrogen atoms were frozen.

\begin{figure}
\includegraphics[angle=0,width=4cm,draft=false]{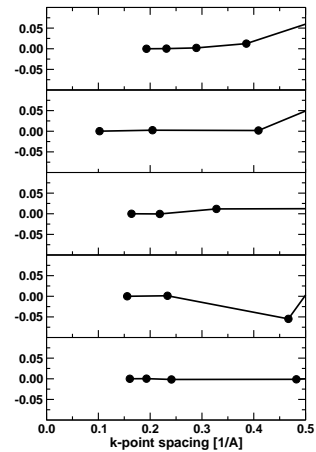}
\caption{K-point convergence: Energy in eV relative to the converged value versus characteristic
k-point spacing.\cite{footnote1} 
From top to bottom: bulk silicon per atom; adsorption
energy per Sr-atom for a coverage of 1/2~ML; bulk silicides
SrSi $Cmcm$, SrSi$_2$ $I4_1/amd$ and SrSi$_2$ $P4_332$ per atom.  Our
surface calculations used a k-point spacing of 0.2\,\AA$^{-1}$ or the closest
commensurate mesh.}
\label{fig:kconv}
\end{figure}

Many of the Sr adsorption structures are metallic which requires a
sufficiently fine grid in k-space. We used an equivalent to eight by
eight points per $(1\times 1)$ surface unit cell.  This value has been
chosen after careful convergence tests for surface structures, bulk
silicon and bulk Sr silicides (Fig.~\ref{fig:kconv}).  In cases where this
k-mesh is incommensurate with the size of the unit cell we selected the
closest, finer commensurate k-mesh.

For metallic systems, the orbital occupations were determined using the
Mermin functional\cite{Mermin} which produces a Fermi-distribution for
the electrons in its ground state. The electron temperature was set to
1000 K. In our case this temperature should not be considered as a
physical temperature but rather  as a broadening scheme for the states
obtained with a discrete set of k-points.  The Mermin functional adds an
entropic term to the total energy, which is approximately canceled by
taking the mean of the total energy $U(T)$ and the Mermin-free energy
$F(T)=U(T)-TS(T)$ as proposed by Gillan:\cite{Gillan89}

\begin{equation}
U(T=0)\approx\frac{1}{2}(F(T)+U(T)).
\end{equation}

In order to express our energies in a comprehensible manner, we report
all energies relative to a set of reference energies.  This set is
defined by bulk silicon and the lowest energy polymorph of SrSi$_2$
$(P4_332)$. The reference energies are listed in
Tab.~\ref{tab:reference}. The reference energy \mbox{$E_0[$Sr$]$} for a
Sr atom, corresponding to the coexistance of bulk silicon and bulk
SrSi$_2$, is extracted from the energy E[SrSi$_2$] of the disilicide
calculated with a $(8\times 8\times 8)$ k-mesh and the reference energy
of bulk silicon $E_0[Si]$ as

\begin{equation}
E_0[\mathrm{Sr}]=E[\mathrm{SrSi}_2]-2E_0[\mathrm{Si}].
\end{equation}

The bulk calculation for silicon was performed in the two atom
unit cell with a ($10\times 10\times 10$) k-mesh and at the experimental
lattice constant of 5.4307~\AA.\cite{CRC} 

\begin{table}
\begin{tabular}{lr}
\hline
\hline
                   & \hspace*{0.3cm} Energy [H] \\ \hline
$E_0$[Si]          &    -4.0036\\
$E_0$[Sr]          &   -31.1441\\
$E_0$[5 layer-Si-slab] &   -21.1140\\
$E_0$[4 layer-Si slab] &   -17.1083\\
\hline
\hline
\end{tabular}
\caption{Reference energies used in this paper without frozen core energy. See text for
details of the calculation.}
\label{tab:reference}
\end{table}

For the surface calculation, we always subtracted the energy of a clean
($4\times2$) silicon surface of the same slab thickness, to account for
slab including hydrogen termination.  For the surfaces with half a
monolayer coverage of silicon we assumed that the corresponding
reservoir for the silicon atoms is a silicon terrace. Hence, the
reference energy for the 4.5 layer silicon slab is the average energy of
a 4-layer slab and a 5-layer slab.  The terrace energy itself does not
enter the limit of an infinitely dilute step density.

In some of our structures the choice of unit cell has an impact on the
dimer buckling. We estimated the energy of a buckling reversal from
the energy difference of a $(2\times2)$ and a $(5\times2)$ supercell.
The cell with an odd number of dimers contains one buckling reversal.
The calculated energy for such a buckling reversal is  0.06~eV.
%
\section{Bulk Silicides}

Before studying the adsorption of Sr on silicon we investigated the bulk
silicides of Sr. The energetics of bulk silicides provide us with the
driving force to go from ordered surfaces structures to silicide grains
on the surface. Our calculations on early transition metals on silicon
indicate that silicide formation is a major problem for layer-by-layer
growth of an oxide.\cite{Foerst03}

The binding characteristics of the bulk silicides provide us with
insight into the favored structural templates which might be anticipated
for the Sr-covered silicon surface.

Sr-silicides are typical Zintl compounds.  According to the Zintl-Klemm
concept,\cite{Zintl39} atoms with an increased number of  electrons form
similar structures as atoms with the correspondingly increased atomic
number.  Consequently, a charge transfer of one electron to silicon will
result in a preferred bonding environment similar to phosphorous with
three or five covalent bonds.  Addition of two electrons will result in
chain-like structures like sulfur. After addition of three electrons one
anticipates formation of dimers and once four electrons are transfered,
isolated ions are expected.  In other words: for every added electron
one covalent bond will be missing.  

Due to the large difference in electronegativity, Sr formally donates its
two valence electrons to the silicon substrate.

\begin{itemize} 
\item In SrSi$_2$
one electron is transfered per silicon atom. Hence three-fold
coordinated silicon networks are formed  as shown in the top two
structures of Fig.~\ref{fig:silicides}.

\begin{figure}
\includegraphics[angle=0,width=8cm,draft=false]{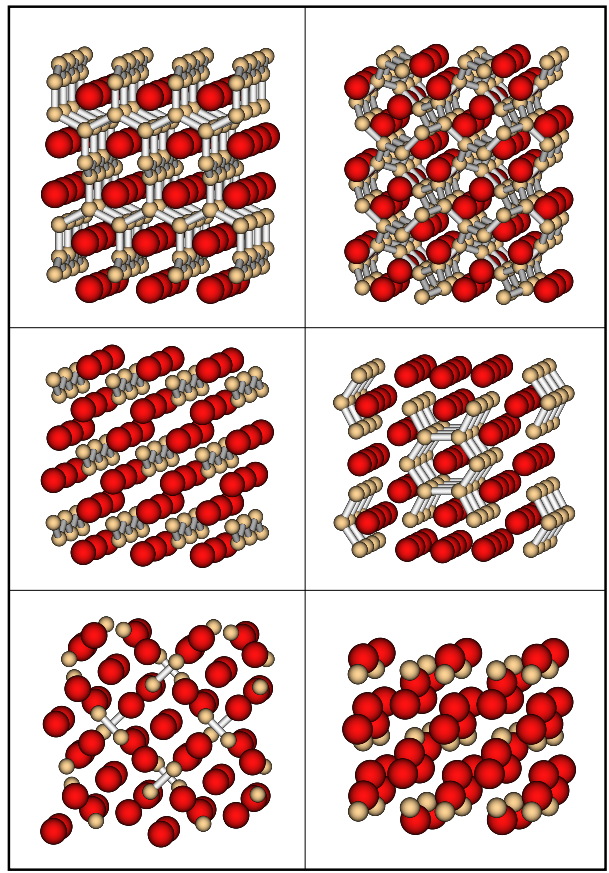}
\caption{Bulk silicide structures. 
Top left: SrSi$_2$ $(I4_1/amd)$\cite{srsi2i4}; 
Top right: SrSi$_2$ $(P4_332)$\cite{srsi2p4}; 
Middle left: SrSi $(Cmcm)$\cite{srsicmcm}; 
Middle right: SrSi $(Immm)$\cite{srsiimmm};
Bottom left: Sr$_5$Si$_3$  $(I4/mcm)$\cite{sr5si3}; 
Bottom right: Sr$_2$Si ($Pnma$)\cite{sr2si}.
The large, dark spheres represent Sr atoms, the smaller, 
light spheres Si atoms. Energies are listed in Tab.~\ref{tab:bulksilicides}.
}
\label{fig:silicides}
\end{figure}

\item Two electrons are transfered in SrSi so that the silicon network
is similar to that of elemental sulfur with two-coordinated silicon
atoms forming Si chains as seen in the middle left panel of
Fig.~\ref{fig:silicides}. The middle right panel shows another
modification of SrSi. The average number of covalent bonds per silicon
atom is, however, still two.

\item In Sr$_5$Si$_3$ there is a charge transfer of 10 electrons to
three silicon atoms, which can be used to form two Si$^{3-}$ ions and
one Si$^{4-}$ ion. The two Si$^{3-}$ combine to form dimers and the
Si$^{4-}$ is no more able to form covalent bonds. Hence we observe an
equal number of Si dimers and single Si ions in the structure of
Sr$_5$Si$_3$ as shown in the lower right panel of Fig.~\ref{fig:silicides}.

\item In Sr$_2$Si four electrons are transfered to each silicon atom. As
a consequence, the Si atoms in the stucture on the bottom right of
Fig.~\ref{fig:silicides} do not form covalent bonds.

\end{itemize}

We find SrSi$_2$ $(P4_332)$ to be the most stable phase of silicides per
Sr atom (Tab.~\ref{tab:bulksilicides}). Therefore we have chosen this
material to define, together with bulk Si, the reference energy for Sr. 

\begin{table}
\begin{tabular}{lr}
\hline
\hline
                    &\hspace*{1cm} E[Sr] [eV]\\
\hline
SrSi$_2$ $(P4_332)$       &  0.00 \\
SrSi$_2$ $(I4_1/amd)$     &  0.01 \\
SrSi $(Cmcm)$             &  0.09 \\
SrSi $(Immm)$             &  0.20 \\
Sr$_5$Si$_3$  $(I4/mcm)$  &  0.39 \\
Sr$_2$Si ($Pnma$)         &  0.45 \\
\hline
\hline
\end{tabular}
\caption{Energies per Sr atom of bulk silicides relative to our reference energies.} 
\label{tab:bulksilicides}
\end{table}

As a side remark, we note that a lower energy of a Sr atom in a bulk
silicide compared to the adsorbed Sr on the surface does not
automatically indicate the formation of silicide grains during growth:
The silicide formation may be suppressed by the strain due to an
epitaxial constraint by the silicon lattice constant. Thus the formation
of silicides is expected to be delayed for thin films, because the bulk
silicides have a large mismatch with the silicon substrate. This
argument does not refer to the thermodynamic equilibrium of large
samples, but it indicates that nucleation of silicide grains will have
to overcome a large barrier.

%
\section{The clean silicon surface} \label{sec:cleansurf}

The clean (001) silicon surface has a $c(4\times2)$ dimer-row
reconstruction. We briefly summarize the driving forces towards this
reconstruction in order to understand the adsorption structures of Sr on
silicon.

An unreconstructed  surface of silicon (001) is terminated
by a square, ($1\times1$) array of atoms. Each silicon atom on the surface is
connected by two bonds to the subsurface. Consequently, there are
two half-occupied dangling bonds on each silicon sticking out of the
surface.

Each pair of surface silicon atoms forms a dimer bond which
saturates one of the two dangling bonds on each atom.  The dimers
arrange in rows. This is the so-called $(2\times1)$ dimer row
reconstruction which results in an energy gain of 0.65~eV per
$(1\times 1)$ unit cell according to our calculations.

In a second reconstruction, both electrons in the dangling bonds
localize on one atom of each dimer, resulting in a dimer buckling. The
buckling is driven by the fact that a sp$^3$ hybridization is favored
for a five-electron species such as the negative Si atom, while a sp$^2$
hybridization is favored for a 3-electron species such as the positive
silicon atom. The sp$^2$ hybridization in turn favors a planar bonding
environment, whereas three-coordinated sp$^3$ bonded atoms form an
umbrella like environment. In the buckled-dimer row reconstruction, the
electrons are localized on the atoms sticking out farthest from the
surface. The dimer buckling can be considered as a Peierl's distortion
which splits the half-filled energy bands resulting from the dimer bonds
into a filled and an empty band, with a band gap in between. The energy
gain due to this distortion is 0.12~eV per $(1\times 1)$ cell.  

The energies quoted here are in reasonable agreement with previously
published LDA pseudo-potential calculations.\cite{Ramstad94} The dimer
reconstruction can be considered to be fairly stable.  Even at 1500~K
only 3\% of the dimer bonds are broken as estimated from the Boltzmann
factor with $\Delta E=0.65+0.12$~eV.

Two neighboring buckled dimer rows interact only weakly. We obtain an
energy difference of 1.2~meV per dimer between the $c(4\times2)$
reconstruction with anti-parallel buckling and the $p(2\times2)$
reconstruction with parallel buckling on neighboring dimer rows. This
indicates that the buckling patterns of different rows are fairly
independent of each other. 

Within a row, however, the buckling of the dimers is coupled in an
anti-correlated manner. This can, at least partly, be explained by the
fact that the lower silicon atom of a dimer pushes the two adjacent
subsurface silicon atoms apart.  For the equivalent silicon atom of the
next dimer, it is therefore favorable to be in the higher, sp$^3$-like
configuration.\cite{Wolkow92}

There has been an intense debate as to whether there is dimer buckling
or not.  STM images reveal a $2\times1$ structure. They exhibit the
buckling only at rather low temperatures and near defects.  The
theoretical predictions depend strongly on the approach chosen (cluster
calculations with configuration interaction or density functional
calculations with periodic boundary conditions).  The most conclusive
results have been produced by Quantum Monte Carlo (QMC)
simulations\cite{SchefflerQMC} indicating that the buckling is present
and density functional calculations just overestimate the energy
difference.

The fact that STM experiments cannot resolve the dimer buckling may be
due to thermal averaging of the two buckled configurations. We believe
that the mechanism is due to the migration of a soliton-like defect in
the anti-correlated buckling pattern of a dimer row.  The calculated
energy for this defect is 0.06~eV (see section~\ref{sec:details}).  Thus
we predict a concentration of one such defect per 11 dimers at room
temperature. Typical tunnel currents in STM experiments are around 1~nA,
which corresponds to six electrons per nanosecond.  

Thus even a soliton migration barrier as large as 0.05~eV would imply
that the buckling changes once during the transfer of a single electron.
These estimates should be taken with caution, since the small energy
difference of 0.06~eV carries a large relative error bar.

From the comparison between DFT and Quantum Monte Carlo calculations one
can deduce an error bar of 0.05~eV per ($1\times1$) unit cell due to
electron correlations.\cite{SchefflerQMC}  We verified that this
correction does not qualitatively affect the findings reported in this
paper.

%
\section{Isolated Sr on silicon}\label{sec:iso}

\begin{figure}
\includegraphics[angle=0,width=8cm,clip=true,draft=false]{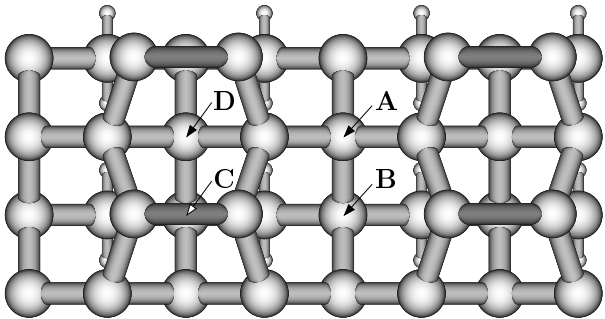}
\caption{Top view of the Si(001) surface and the four high
symmetry positions spanning the surface irreducible $(2\times1)$ unit cell.
The dimer buckling is not shown. The energies are listed in 
Tab.~\ref{tab:isolatedenergies}.}
\label{fig:surfirrzone}
\end{figure}

In order to determine the low-coverage limit of Sr-adsorption we
investigated the energy as a function of the lateral position of a Sr
ad-atom on the surface.  The total energy as a function of the lateral
position of Sr is obtained by constraining the lateral movement of the
Sr-atom relative to the rigid back plane of the slab.  The calculations
were performed in a $(4\times4)$ surface super cell.  We considered the
high symmetry points of the (2$\times 1$) surface shown in
Fig.~\ref{fig:surfirrzone} and the mid-point between the local minima.
The energies of the high symmetry positions are given in
Table~\ref{tab:isolatedenergies}.

\begin{table}
\begin{tabular}{cr}
\hline\hline
Position & $\hspace*{1.0cm}\Delta E$ [eV]\\
\hline
A      & 0.00\\
B      & 0.55\\
C      & 0.75\\
D      & 0.29\\ 
A--D midpoint & 0.61 \\ \hline\hline
\end{tabular}
\caption{Relative energies of isolated Sr on the Si(001) surface at the
  high-symmetry points of the surface irreducible unit cell as well as
  the A--D midpoint.  The
  labels refer to Fig~\ref{fig:surfirrzone}.}
\label{tab:isolatedenergies}
\end{table}

Sr has the global minimum at position \textbf{A} as defined in
Fig.~\ref{fig:surfirrzone}. Sr is located in the trench between the
dimer rows and in the center of 4 surrounding dimers. The Sr atoms are
slightly elevated above the plane of the surface dimers. 

A metastable position, \textbf{D} of Fig.~\ref{fig:surfirrzone}, is
located in between two dimers on top of a dimer row. It is 0.29~eV
higher in energy than the global minimum. We will see that the
structures \textbf{A} and \textbf{D} are repeating motifs in a range of
different adsorption structures.

The diffusion of Sr on the silicon surface proceeds about equally fast 
parallel and perpendicular to the dimer rows, with a slight preference
for the parallel direction.  The diffusion barrier along the valley is
equal to the energy difference between sites \textbf{A} and \textbf{B},
namely 0.55~eV, the one across the row is 0.61~eV and is estimated by
the midpoint between the sites \textbf{A} and \textbf{D}

It should be noted that in our analysis we ignored the reduced symmetry
due to dimer buckling.  As a result, different versions of the high
symmetry points quoted here exist with slightly different energies. For
the structure A we found two versions which differ in energy by 0.15~eV.
In these cases the lowest energy structure has been chosen.

In contrast to Sr, there has been a lot of work related to isolated Ba
atoms adsorbed on Si(001).\cite{Wang99,Yao99} Our results are in line
with previous calculations for isolated Ba on Si(001).\cite{Wang99} We
believe that both atoms behave in a similar fashion. Ba has been found
mostly on sites \textbf{A} in the trenches, but also on sites \textbf{D}
on top of the dimer rows.  The main difference between Ba and Sr lies in
the energy difference between the two metastable sites \textbf{A} and
\textbf{D}. For Ba the difference is 0.88~eV~\cite{Wang99} which is
substantially larger than the 0.29~eV for Sr.

The chemical binding can be well understood in an ionic picture as
suggested by the chemical binding of the silicides. The two electrons of
the Sr atom are donated into an unoccupied dangling bond of a Si dimer.
Interestingly we find this electron pair to be localized at a single
dimer. This is evident from the dimer buckling, which vanishes when both
dangling bonds are occupied.

\begin{figure}
\includegraphics[angle=0,width=6cm,draft=false]{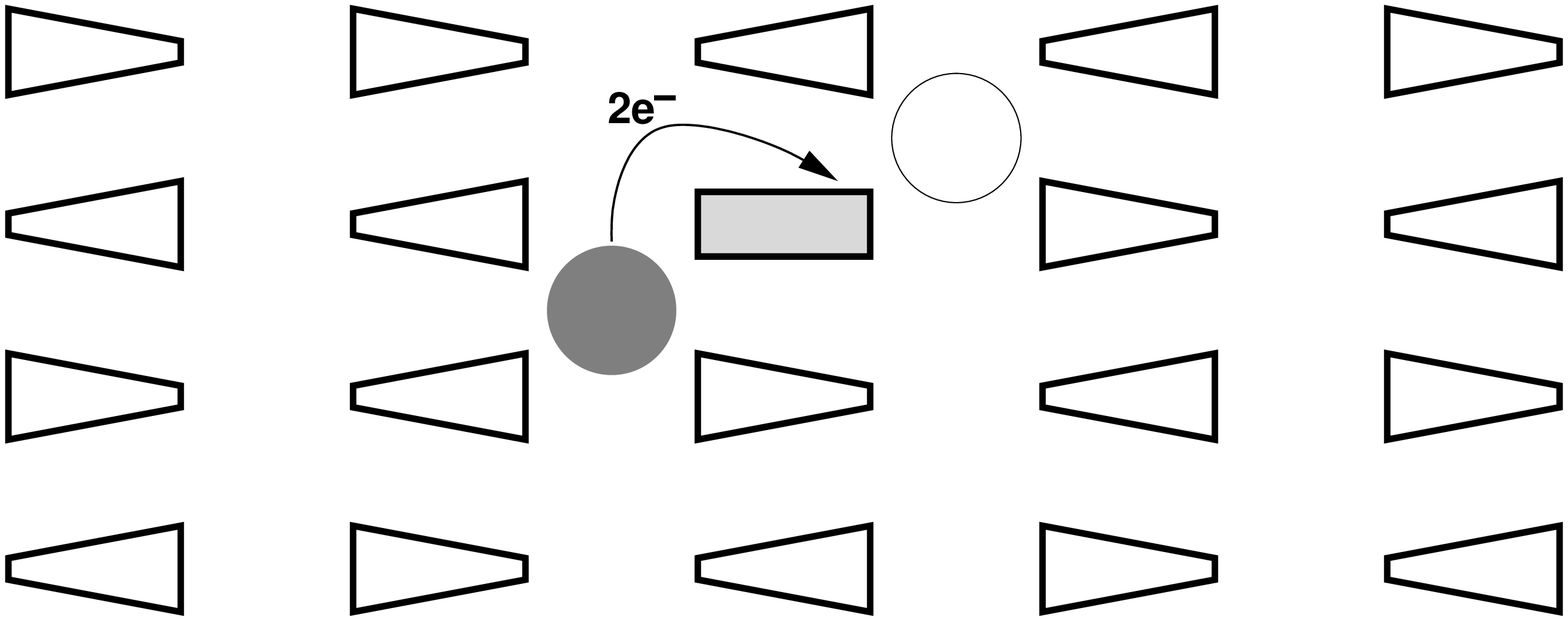}
\caption{Schematic representation of the isolated Sr ad-atom at
position {\bf A}. The filled circle represents the Sr ad-atom, the rectangle
represents a filled and therefore unbuckled Si dimer. The triangles
represent buckled dimers.  The flat side of a buckled dimer
indicates the upper Si atom with a filled dangling bond, whereas the
pointed side indicates the lower Si atom with the empty dangling
bond. The charge transfer from the Sr ad-atom to one of the surrounding
dimers is indicated by the arrow, the preferred adsorption site (see
section~\ref{sec1.5}) in the
neighboring valley by the open circle.}
\label{fig:posa}
\end{figure}

The Sr atom experiences an additional electrostatic stabilization from
the remaining three buckled dimers next to it.  They are buckled such
that the negative, and therefore raised, silicon atoms are located next
to the Sr atom. The local configuration is shown in Fig.~\ref{fig:posa}.
This arrangement significantly affects the buckling of the two dimer
rows adjacent to the Sr atom: (1) the buckling gets pinned and is
therefore observed in STM images in the vicinity of a Sr atom whereas it
is thermally averaged out on the bare surface;  (2) the dimer buckling
within one row is reversed as already pointed out by Wang et
al.\cite{Wang99} This becomes apparent by looking at the row left of the
Sr atom in Fig.~\ref{fig:posa}.

From the static structure shown in Fig.~\ref{fig:posa} it is not evident
why \textit{both} dimer rows contain a buckling reversal as observed in
Fig.~1\,b) of Yao et al.\cite{Yao99} We attribute the experimental
observation to a dynamical effect: One electron pair can rapidly migrate
from the filled dimer to one of the three buckled dimers next to the Sr
ad-atom. If this fluxional motion occurs on a time scale faster than the
time scale for a buckling reversal of the entire chain, the buckling
will appear pinned in both dimer rows.

In our super cell with an even number of dimers in a row, every reversal
of the buckling must be compensated by a second one, thus artificially
destabilizing the site \textbf{A}.  This adds an uncertainty of up to
0.12~eV to all energies for the isolated Sr.  Even taking this
uncertainty into account, isolated Sr ad-atoms do not form a
thermodynamically stable phase at any coverage as will be demonstrated
in the next section.

%
\section{Chain structures at dilute coverages}\label{sec1.5}
The Sr-atoms on the surface tend to arrange in chains, as seen in the
STM experiments.\cite{Bakhtizin96-1,Bakhtizin96-2} Similar results
have been obtained for Ba on Si(001).\cite{Hu00} 

Our calculations predict random, single chain structures up to a
coverage of 1/6~ML. Between 1/6~ML and 1/4~ML we find condensed, single
and double chain structures. Above 1/4~ML the multiple chain structures
convert into disordered arrays of double vacancies as will be discussed
below.  We investigated chain structures with coverages of 1/16, 1/10,
1/8, 1/6, 1/4, 3/10, 1/3, 5/14 and 6/16~ML.  We find an energy gain of
0.3 to 0.4~eV per Sr atom when single chains are formed from isolated Sr
ad-atoms on \textbf{A}-sites. 

\begin{figure}
\includegraphics[angle=0,width=8cm,draft=false]{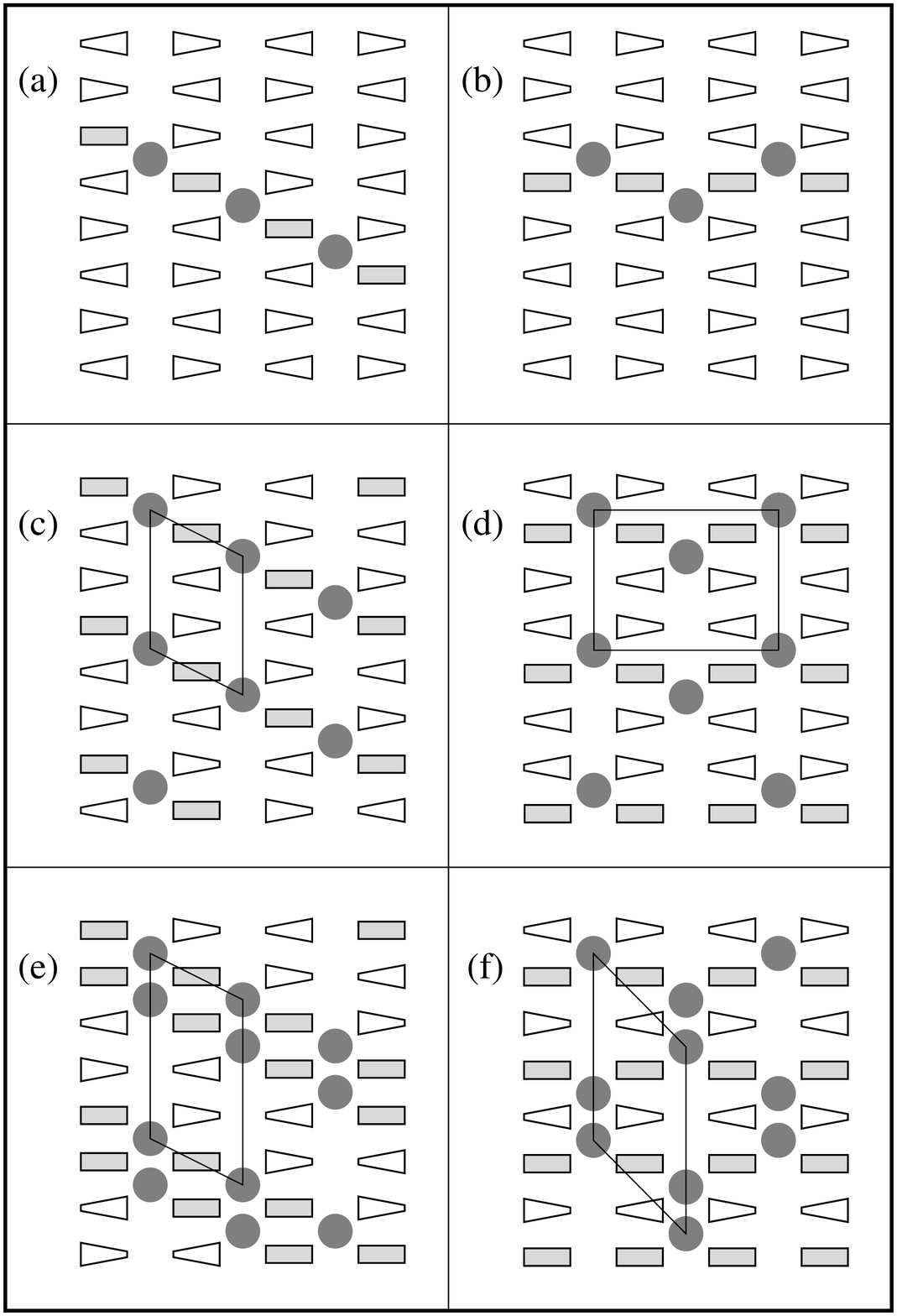}
\caption{Schematic representation of a set of chain structures.
Isolated chains are shown in the top row. The ordered single chain
structures at 1/6 ML are shown in the middle row. The bottom row shows
the ordered structures at 1/4 ML. The left hand side shows straight
(diagonal) chains running at an angle of 60~\% to the dimer rows, while
the right column shows zig-zag chains perpendicular to the dimer rows.
See Fig.~\ref{fig:posa} for a description of structural elements. The
surface unit cells are outlined. The energies are listed in
Tab.~\ref{tab:ene}.} 
\label{fig:schematic1614} 
\end{figure}

\begin{table}
\begin{center}
\begin{tabular}{llr}
\hline\hline
Sr coverage \hspace*{0.5cm}& Figure number \hspace*{0.5cm} & Energy/Sr [eV] \\ 
\hline
1/$\infty$      & \ref{fig:schematic1614}(a)    & just schematic \\
1/$\infty$      & \ref{fig:schematic1614}(b)    & just schematic \\
1/6      & \ref{fig:schematic1614}(c)    & -1.15\\ 
1/6      & \ref{fig:schematic1614}(d)    & -1.15\\ 
1/4      & \ref{fig:schematic1614}(e)    & -1.10\\ 
1/4      & \ref{fig:schematic1614}(f)    & -1.06 \\ 
1/4      & \ref{fig:mckee}~bottom           & -0.30 \\ 
1/4      & \ref{fig:mckee}~top              & 0.44 \\ 
1/2      & \ref{fig:strchalfml}          & -0.92 \\ 
2/3      & \ref{fig:2by3and1ML}~left        & -0.74 \\ 
1        & \ref{fig:strc1MLprinciple}~left  & 0.04 \\ 
1        & \ref{fig:strc1MLprinciple}~right & 0.08 \\ 
1        & \ref{fig:2by3and1ML}~right       & -0.12 \\ 
4/3      & \ref{fig:4by3ML}                 & 0.28 \\ 
\hline\hline
\end{tabular}
\end{center}
\caption{Energies per Sr ad-atom relative to our reference energies for all
structures graphically represented in the figures unless not already
tabulated elsewhere in the paper.}
\label{tab:ene}
\end{table}

The chain formation is driven by the electrostatic attraction between
the positively charged Sr-ions, located at \textbf{A}-sites, and the
negatively charged dimers: An isolated Sr ion located at an
\textbf{A}-site, with the lone pairs of the four neighboring dimers
pointing towards it, donates its two valence electrons into a silicon
dangling bond adjacent to a neighboring valley.  Thus it offers a
preferred binding site for a Sr atom in that valley, namely next to this
filled dangling bond as seen in Fig.~\ref{fig:posa}. This second atom in
turn donates its electron pair to the dimer row which does not already
contain a negatively charged dimer and all four surrounding dimers will
rearrange to point their lone pairs towards the new ad-atom.  This
process continues to form chains of Sr atoms.  The filled dimers are
clearly identified by the absence of any buckling. 

Two nearest Sr ions of one chain are displaced parallel to the dimer row
by one lattice constant in order to position the filled dimer inbetween.
Since the favorable Sr positions are staggered with respect to the
dimers, the chains run at an angle of 30~degree relative to the
direction of a dimer or 60 degree with respect to the dimer rows. The
small energy difference of 0.02~eV per Sr ion between diagonal and
zig-zag chains at e.g.  1/8th ML cannot be considered as a hard number
due to the systematic errors of DFT calculations. It does, however,
indicate that the energy cost for changing the direction of such a chain
is negligible so that these chains may meander on the surface.

As for the isolated Sr ad-atom, an additional stabilization occurs due
to the dimer buckling of the surrounding silicon dimers. The negatively
charged, raised part of an adjacent dimer is located next to the Sr-ion,
stabilized by electrostatic and covalent interactions. Reversing the
buckling of one of the dimers next to a Sr ion raises the energy by
0.38~eV.\cite{clemens} This ordering induces a freezing of the dimer
buckling, which reaches far out into the clean silicon surface, as can
be clearly seen in the STM images by Hu et al.\cite{Hu00} 

At first sight one might think that there is a second preferred binding
site in Fig.~\ref{fig:posa}, right next to the initial ad-atom on the
\textbf{A} site below the open circle.  This configuration is, however,
only possible for a pair of Sr atoms. A further continuation of such a
chain perpendicular to the dimer row will make it impossible to
rearrange dimers in a way that only filled dangling bonds are oriented
towards the ad-atoms. Such chains are therefore destabilized with
respect to diagonal or zig-zag ones.

A favorable registry between two Sr-chains is obtained if the Sr atoms
are either in contact or separated by an even number of vacant
\textbf{A} sites along each valley of the Si-surface.  This follows from
a simple building principle which is an extension of what is already
known from isolated chains:

\begin{itemize}

\item Each Sr atom is electrostatically stabilized by four negatively
charged silicon atoms located next to it. Negative silicon atoms have
two electrons in their dangling bond and are in a raised, $sp^3$-like
bonding configuration. 

Violation of this rule raises the energy by 0.38~eV per
empty dangling bond next to the Sr atom. At this level of abstraction we
do not distinguish between a negatively charged Si atom of a buckled and
an unbuckled dimer.

\item There are no reversals of the dimer buckling in the Sr-free
regions on the surface. A buckling reversal increases the energy by
0.06~eV. This is a consequence of the anti-correlated coupling of the
dimer buckling within one row (see section~\ref{sec:cleansurf}). 

\end{itemize}

When the chains approach the shortest possible distance before they
collapse into double chains, we obtain a partially ordered structure at
1/6 ML as shown in Fig.~\ref{fig:schematic1614} (c) and (d). It should
be noted that our calculations indicate that also the condensed chains
at 1/6~ML change their directions frequently, even though synchronized
with the neighboring chains running in parallel.  

As the coverage increases, Sr atoms arrange themselves into double
chains as shown in Fig.~\ref{fig:schematic1614}(e), resulting in a
partially ordered surface structure at 1/4 ML.  The "1$\times$2" areas
in Fig.~4 of Ojima et al.\cite{Ojima02} can be explained by double
chains at a 1/4~ML.  Their interpretation that buckled Ba dimers at
1/2~ML coverage are responsible for this "wavy structure" cannot be
supported by our calculations.

If we continue this building principle beyond 1/4 ML we will obtain
sequences of triple, quadruple, etc., chains of Sr atoms separated by
double vacancies. However, the positions of the double vacancies of
different valleys are then only weakly correlated.  Double vacancies of
neighboring valleys can arrange themselves almost arbitrarily except
that they do not line up perpendicular to the dimer rows. This implies a
new building principle of double vacancies that do not necessarily
arrange in chains.  Note that a multiple chain structure of Sr atoms can
also be interpreted as a chain structure of double Sr vacancies.  This
building principle can already be observed in
Fig.~\ref{fig:schematic1614}(e) and (f) for 1/4~ML, which show that
double vacancies in neighboring valleys can assume three out of four
relative positions. While the energy difference between the structures
in Fig.~\ref{fig:schematic1614}(e) and (f) is as small as 0.04~eV per Sr
atom, it decreases further for the analogous structural patterns for
wider Sr-chains.  For coverages close to one-half ML, when the
concentration of double vacancies is dilute, we therefore expect a
nearly random arrangement of double vacancies instead of multiple
chains.

The reciprocal lattice vectors for the diagonal chain structures are
$(\frac{1}{2},0)$ and $(\pm \frac{1}{2n}, \frac{1}{n})$, where $n$ is
the periodicity of the the real space unit cell (compare
Figs.~\ref{fig:schematic1614}(c) and (e)) along the dimer row direction.
It should be noted that there is some structural disorder due to
frequent changes in the chain direction.  For zig-zag chains the
corresponding reciprocal lattice is spanned by the vectors
$(\frac{1}{4},0)$ and $(0, \frac{1}{n})$. The actual diffraction pattern
observed in experiment will contain a mixture of both reciprocal lattices.

\begin{figure}
\includegraphics[angle=0,width=8cm,draft=false,clip=true]{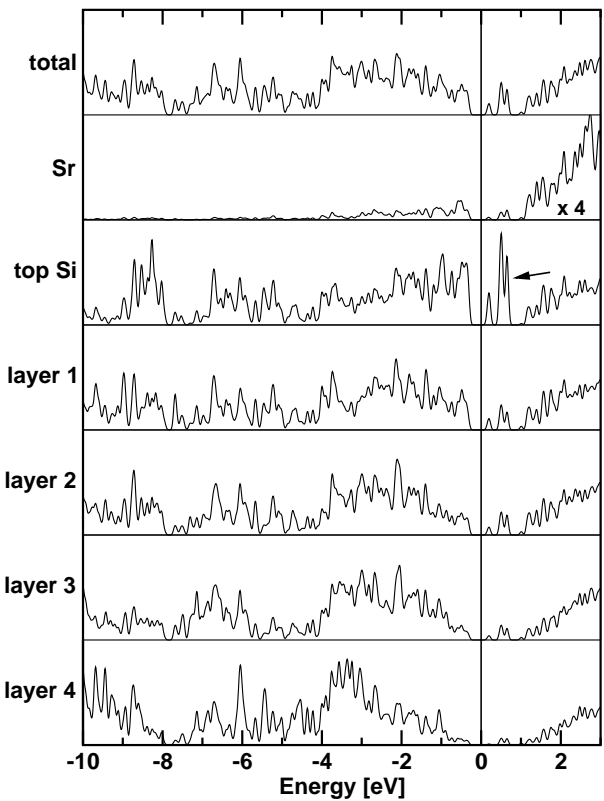}
\caption{Layer-projected density of states of the single chain
  structure at 1/6~ML. Values are divided by the number of atoms
  in the layer. The vertical line at 0~eV indicates the Fermi level, the
  arrow points at the characteristic surface band in the bandgap of
  silicon.
  }
\label{fig:dos_1by6ML}
\end{figure}

\begin{figure}
\includegraphics[angle=0,width=8cm,draft=false,clip=true]{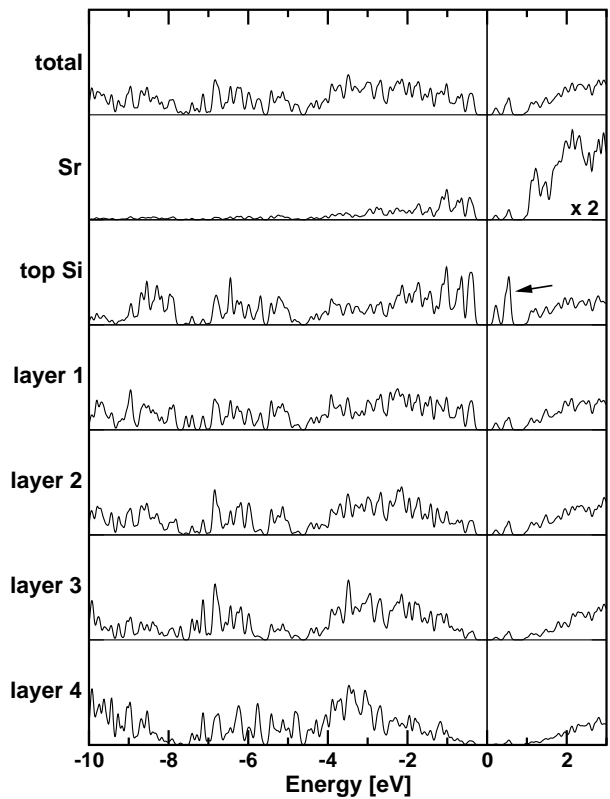}
\caption{Layer-projected density of states of the double chain
  structure at 1/4~ML. Values are divided by the number of
  atoms in the layer. The vertical line at 0~eV indicates the Fermi
  level, the arrow points at the characteristic band states in the bandgap of silicon.}
\label{fig:dos_1by4ML}
\end{figure}

\begin{figure}
\includegraphics[angle=0,width=8cm,draft=false,clip=true]{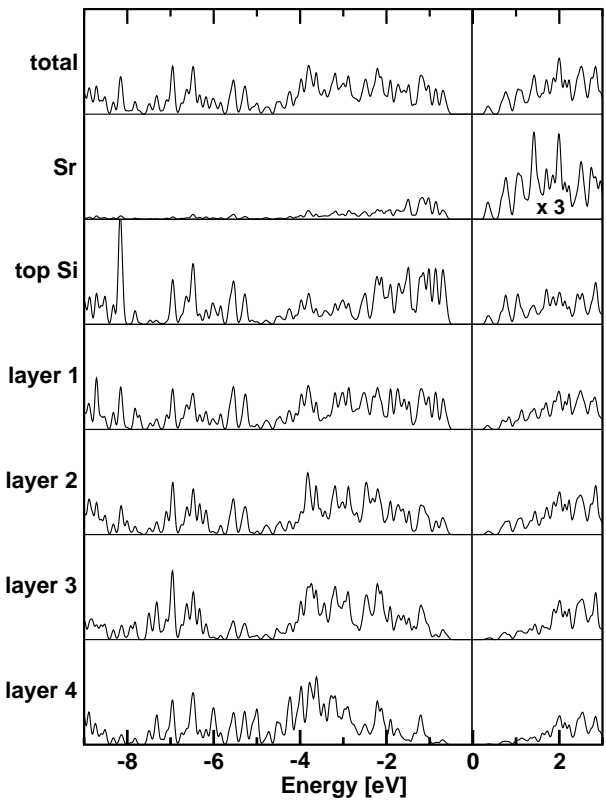}
\caption{Layer-projected density of states of the $(2\times1)$
  structure at 1/2~ML. Values are divided by the number of
  atoms in the layer. The vertical line at 0~eV indicates the Fermi
  level.}
\label{fig:dos_1by2ML}
\end{figure}

The density of states of the single chain structure at 1/6~ML
is shown in Fig.~\ref{fig:dos_1by6ML}. The states on the Sr atom, which
appear in the valence band, can be attributed to the tails of the Si
dangling bonds, which hybridize with the Sr-$s$ orbital.  We observe
states in the Si band gap, which are assigned to the empty dangling
bonds on the buckled dimers.  These states actually form a single band
that is separated from the valence and conduction bands. The fact that
they appear as individual states is an artifact of our discrete sampling
of the Brillouin zone.  This band pins the Fermi level in the lower part
of the band gap.  This feature remains nearly unchanged in the density
of states of the double chain structure at 1/4~ML
(Fig.~\ref{fig:dos_1by4ML}).  It disappears, however, with the absence
of the half-occupied dimers at 1/2~ML as seen in
Fig.~\ref{fig:dos_1by2ML}.  Thus the gap states remain approximately in
their position as the coverage increases, but the density of states is
reduced. Hence the Fermi-level will remain pinned in the lower part of
the silicon band gap up to a coverage of 1/2~ML. At this point
the Fermi-level becomes unpinned. In the following we will see that for
coverages above 1/2~ML states from the conduction band are
pulled into the band gap of silicon, and pin the Fermi level in the
upper part of the band gap.

This finding explains the discontinuity of the band-bending as observed
by Herrera-Gomez et al.\cite{Herrera01} Their XPS studies show that the
Fermi level shifts up by almost 0.5~eV when the coverage is increased
from below 1/2~ML to above.  However, it should be noted that such a
discontinuity in Fermi-level pinning is not specific to detailed
structures:  also higher energy structures exhibit a similar behavior.

%
\section{The $\mathbf{(2\times 1)}$ reconstruction at 1/2 ML}

\begin{figure}
\centering
\includegraphics[angle=0,width=8cm,draft=false]{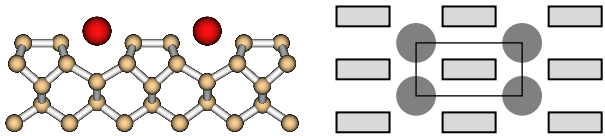}
\caption{The (2$\times$1) reconstructed surface at 1/2~ML coverage.
The energy is listed in Tab.~\ref{tab:ene}.}
\label{fig:strchalfml}
\end{figure}

At a coverage of 1/2~ML all dangling bonds of the surface dimers are
fully occupied (Fig.~\ref{fig:strchalfml}). It can be considered as  the
canonical Sr-covered Si surface.  It is the only Sr-covered surface
structure without states in the band gap of silicon. This structure is
"isoelectronic" to an hydrogen terminated silicon surface and is
therefore expected to be comparably inert.  The increased resistance to
oxidation has already been observed experimentally.\cite{Liang01} A
$(2\times 1)$ reconstruction at 1/2~ML has already been reported by
several authors.\cite{Fan90,Fan91,Urano96,Herrera01,Lettieri02}

This structural template has been proposed to be the basic building
block of the interface between silicon and SrTiO$_3$.\cite{Foerst03-1}

\section{From 1/2 ML  to 1 ML}
%

\begin{figure}
\includegraphics[angle=0,width=8cm,draft=false]{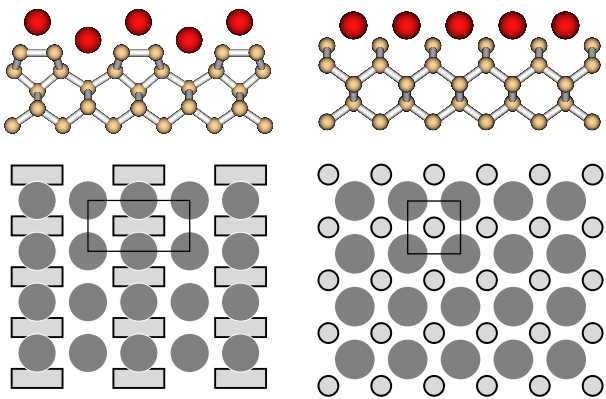}\\
\caption{Building blocks for the surface structure with 1ML coverage
having a $(2\times1)$ (left) and a $(1\times1)$ (right) reconstruction.
The outlined circles represent single Si atoms in the first layer. 
The energies are listed in Tab.~\ref{tab:ene}.}
\label{fig:strc1MLprinciple}
\end{figure}

For coverages between 1/2 ML and 1 ML, we find structures built up from
three structural templates: 
\begin{enumerate}
\item the $(2\times 1)$ reconstruction at 1/2~ML (Fig.~\ref{fig:strchalfml})
\item the $(2\times 1)$ reconstruction at 1~ML
(Fig.~\ref{fig:strc1MLprinciple} left)
\item the $(1\times 1)$ reconstruction at 1~ML
(Fig.~\ref{fig:strc1MLprinciple} right)
\end{enumerate}

When we increase the coverage above 1/2 ML, the additional atoms occupy
the \textbf{D}-sites, since all \textbf{A}-sites are already occupied.
When all \textbf{A}-sites and all \textbf{D}-sites are occupied, as
shown in the left panel of Fig~\ref{fig:strc1MLprinciple}, the coverage
is that of 1 ML.  At this coverage, two electrons are transferred to
each silicon atom instead of only one as in case of the 1/2~ML. These
electrons can fill the dimer antibonding states and thus break up the
dimer bond. When all dimer bonds are broken at 1~ML we obtain a
$(1\times1)$ reconstructed silicon surface with a Sr-ion above the
center of each square of silicon atoms (Fig.~\ref{fig:strc1MLprinciple}
right). This structure is, however, never realized in its pure form due
to the large strain in the top layer.  An indication for the strain is
the difference between  the spacing of Sr atoms in bulk SrSi
(Fig.~\ref{fig:silicides}) and that of this hypothetical surface
structure. The former is larger by 25~\%.  Nevertheless, this pattern is
found as a building block in a number of low-energy structures between
coverages of 0.5~ML and 1~ML. The $(1\times1)$ reconstruction at 1~ML is
unfavorable by 0.04~eV per Sr atom compared to the corresponding
$(2\times1)$ structure.

By combining the $(2\times1)$ structure at 1/2 ML
(Fig.~\ref{fig:strchalfml}) and the $(1\times1)$ structure at 1 ML
(right panel of Fig.~\ref{fig:strc1MLprinciple}), a series of structures
with increasing coverage and periodicity can be formed: dimer rows are
separated by stripes of the ($1\times1$)/1~ML structure with increasing
width.  This leads to a series of $(n\times1)$ structures at a coverage
of $\frac{n-1}{n}$ ML.  We have investigated these structures from $n=3$
to $n=6$.

\begin{figure}
\includegraphics[angle=0,width=8cm,draft=false]{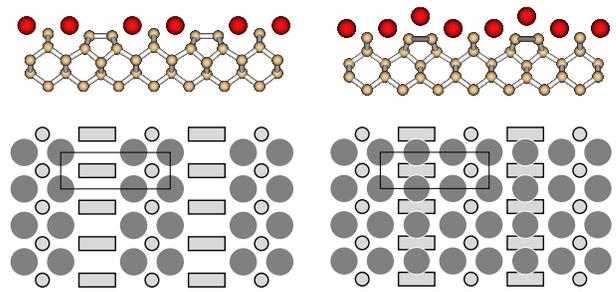}
\caption{The $(3\times1)$ structures for coverages 2/3~ML
  (left) and 1~ML (right). The energies are listed in
  Tab.~\ref{tab:ene}.}
\label{fig:2by3and1ML}
\end{figure}

The first structure in the series is the $3\times1$ reconstructed
surface at a coverage of 2/3~ML shown in the left panel of
Fig.~\ref{fig:2by3and1ML}. It consists of dimer rows separated
by a stripe of two Sr atoms in the ($1\times1$)/1~ML configuration.
According to our calculations this structure is present as a
distinct phase between 0.5~ML and 1~ML. 

In order to form this structure the dimer row pattern needs to
reconstruct. This process is facilitated by the additional electrons
in the conduction band which weaken the dimer bonds:  beyond
a coverage of 1/2 ML Sr is likely to act as a catalyzer for dimer bond
rearrangement.

\begin{figure}
\includegraphics[angle=0,width=8cm,draft=false,clip=true]{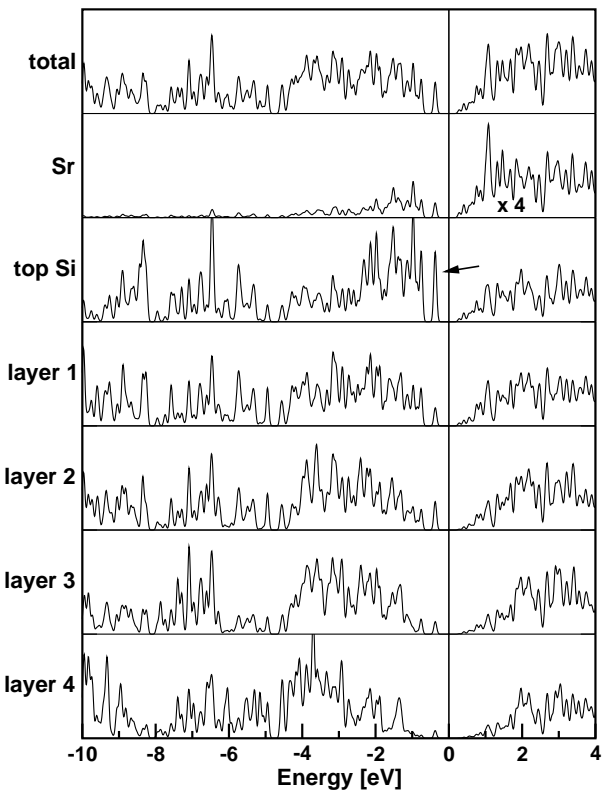}
\caption{Layer projected density of states for $(3\times1)$ surface at
  2/3~ML coverage. Values are divided by the number of atoms
  in the layer.  Note that the characteristic peak just below the
  Fermi level is not a single state, but the top of a surface band
  extending into the continuum of the valence band states. The vertical
  line at 0~eV indicates the Fermi level.}
\label{fig:dos2/3ml}
\end{figure}

The density of states for this $(3\times1)$ structure
(Fig.~\ref{fig:dos2/3ml}) exhibits a surface band that ranges from the
continuum of the valence band of bulk silicon into its bandgap. The
surface band is localized on the dangling bonds of the silicon dimers
and the undimerized silicon atoms.  The Fermi level is pinned between
this band and the conduction band of silicon (see also discussion in
section~\ref{sec1.5}). 

For coverages greater than 2/3~ML, that is $n> 3$, the structures can be
interpreted as stripes with coverage of 1~ML with a $(1\times 1)$
reconstruction separated by dimer rows without Sr, as discussed above.
As the stripes with $(1\times 1)$/1~ML increase in width, they build up
strain, which can be released by forming dimer rows with Sr atoms on top
(Fig.~\ref{fig:strc1MLprinciple} left).  This corresponds to a
transition between 
 both 1~ML structures shown in
Fig~\ref{fig:strc1MLprinciple}.  Since the energy difference between
these structural variants is smaller than 0.04~eV per Sr atom, it is
likely that they do not form distinct phases but transform into each
other in a fluxional fashion.

At a coverage of 1~ML we find a series of structures built from the two
structural templates in Fig.~\ref{fig:strc1MLprinciple}. In our
calculations the $(3\times1)$ (Fig.~\ref{fig:2by3and1ML}) and
$(4\times1)$ reconstructions are most stable and, within the theoretical
error bar, degenerate. There is a slow increase in energy towards the
$(5\times1)$ and $(6\times1)$ reconstructions which are less than
0.11~eV per Sr atom higher in energy.

%
\section{Beyond 1 ML}

\begin{figure}
\includegraphics[angle=0,width=8cm,draft=false]{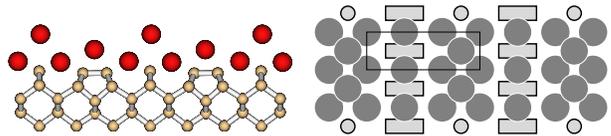}
\caption{4/3~ML coverage showing a 3$\times$1 silicide
layer. The energy is listed in Tab.~\ref{tab:ene}.}
\label{fig:4by3ML}
\end{figure}

Beyond 1~ML additional Sr atoms deposit on top of the silicon atoms in
the $(1\times1)$ stripes.  The first commensurate structure is the
$(3\times1)$ structure at a coverage of $4/3$ ML shown in
Fig.~\ref{fig:4by3ML}.

Fan et al. have observed a $(3\times1)$ reconstruction at a coverage
of 1.3~ML, Bakhtizin et al.\cite{Bakhtizin96-1, Bakhtizin96-2}
have published STM images with a $(3\times1)$ periodicity for a
coverage of 1.2~ML. There, alternating bright
and dark stripes have been observed, which we attribute to the Sr
atoms on top of the dimer rows (darker stripes) and those on top of the
of the two-fold coordinated Si atoms (more prominent stripes).
%
\section{The $\mathbf{(3\times 2)}$ structure at 1/3~ML}

Diffraction and STM studies identify a $(3\times 2)$ reconstruction at 
1/3~ML\cite{Fan90,Fan91,Bakhtizin96-1,Bakhtizin96-2,Hu99,Hu00,Herrera00,Liang01,Ojima01,Ojima02}.
The diffraction studies (LEED, RHEED) did not distinguish between the
orientations parallel and perpendicular to the dimer rows as they
average over multiple terraces.  Most previous studies assumed that the
$3\times$ direction of the $(3\times 2)$ surface unit cell is orientated
parallel to the dimer
rows.\cite{Bakhtizin96-1,Bakhtizin96-2,Herrera00,Droopad03}  However,
the STM images of Hu et al.\cite{Hu00} (Fig.~5) and Ojima et
al.\cite{Ojima02} (Fig.~4) clearly show, that the $3\times$ axis is
orthogonal to the dimer row direction.  This is particularly evident
from the images showing the phase boundaries between the $(3\times 2)$
reconstructed domains and chain structures. This observation implies
that the dimer row pattern is disrupted.

Our lowest energy structures for this coverage are variants of the
quadruple chain as described in section \ref{sec1.5}.  We have been
unable to determine a thermodynamically stable structure with a
$(3\times 2)$ diffraction pattern at 1/3~ML.

Hu et al.~\cite{Hu00} suggested two $(3\times 2)$ structures for 1/6 and
1/3~ML. We simulated both of them and found them energetically
unfavorable compared to the corresponding chain structures by more than
0.61~eV per Sr atom.  Despite various other attempts, we
failed to arrive at a thermodynamically stable surface structure with
this reconstruction.

This failure may be attributed to our inability to scan the entire phase
space.  However, one should also consider the possibility of
co-adsorption of other elements such as hydrogen or oxygen, which may
help to tie up the dangling bonds created by disrupting the dimer row
pattern. These effects have not been considered in the present study.

For the sake of completeness we also studied the model for the 
1/3~ML coverage with the $3\times$ direction parallel to the
dimer row.  It was lower in energy than the structures suggested by Hu
et al.~\cite{Hu00}  Nevertheless it turned out 0.23~eV per Sr atom higher in
energy than the quadruple chain structure.

%
\section{SrSi$_2$ surface layer}
\begin{figure}
\centering
\includegraphics[angle=0,width=6cm,draft=false]{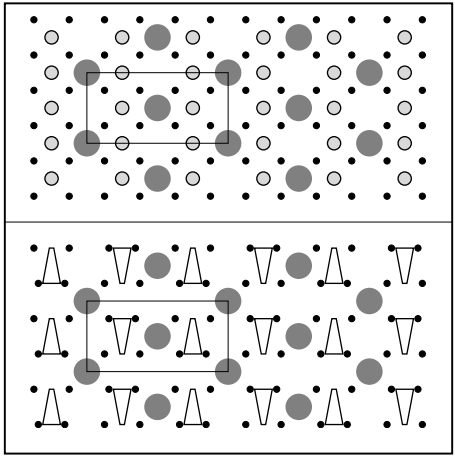}
\caption{Silicide structures with a $c(4\times 2)$ reconstruction at
1/4~ML:
The 1/4 ML $c(4\times2)$ structure as proposed by McKee et
al.~\cite{McKee98} (top)
and the lowest energy structure we find for that chemical composition
(bottom).
The large dark circles correspond to Sr, the smaller, outlined ones to Si in the
silicide layer (only in the top figure) and the small black dots to Si atoms in the
first full layer. In the lower graph the Si atoms of the silicide layer have formed
dimers indicated by the triangles. The energies are listed in
Tab.~\ref{tab:ene}.}
\label{fig:mckee}
\end{figure}

So far, we have discussed structures on a stoichiometric Si-surface.
The atomic model for the interface between Si and SrTiO$_3$ by
McKee et al.,\cite{McKee98} which can be seen in the top panel of
Fig.~\ref{fig:mckee}, has inspired us to also investigate
reconstructions with only 1/2~ML of silicon in the surface layer.  Such
a structure can in principle be formed by the migration of Si atoms or
dimers from step edges onto the terraces. For a clean surface this
process is clearly not energetically favorable. However, it cannot be
excluded a priori that the presence of Sr stabilizes a surface with half
a monolayer of silicon.

We started from the structure proposed by McKee et al.\cite{McKee98}
which consists of a quarter monolayer of Sr and half a monolayer of Si.
In this structure the Sr atoms occupy every second A-site in the
valleys, while the Si atoms deposit on top of the center of four
subsurface Si atoms.  We restricted our search to the $c(4\times 2)$
periodicity reported by McKee et al.

We find a number of metastable structures, the most stable one
(Fig.~\ref{fig:mckee} bottom)  differs substantially from the proposal
by McKee et al.  The half monolayer of silicon on the surface combines
into dimers, a behavior already known from Si ad-atoms on
Si(001).\cite{Smith96}  The Sr atoms occupy positions in the center of
four such dimers.

The energy of this structure is, however, still higher by 0.80~eV per Sr
ad-atom than our lowest energy structure at this coverage, namely double
chains of Sr atoms.  The energy was evaluated relative to the average of
a 4-layer and a 5-layer slab, representing a terrace. This energy thus
describes the process of adsorption of Sr and the decomposition of two
terraces into a single terrace with an additional half-monolayer of
silicon.

We consider the difference in formation energy of 0.80~eV per Sr atom,
relative to our lowest energy structure at this coverage, as too large for
this structure to be physically relevant.

\section{Phase diagram}

\begin{figure}
\includegraphics[angle=0,width=8cm,draft=false]{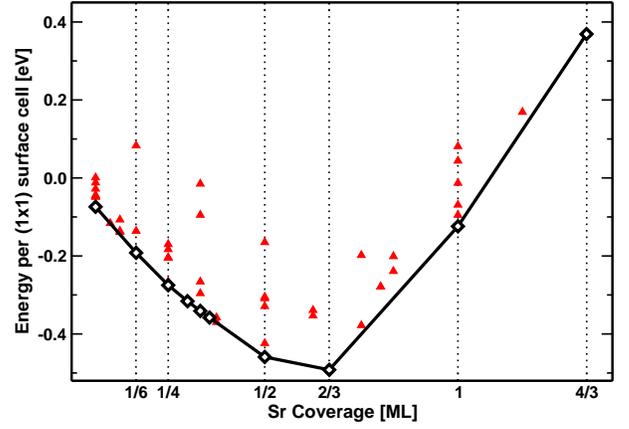}
\caption{The adsorption energy\cite{clemens1} per $(1\times 1)$ unit cell 
  as a function of Sr coverage.  The open diamonds
  represent thermodynamically accessible structures, the triangles
  correspond to metastable structures.}
\label{fig:relenergyvscov}
\end{figure}

We now investigate which of the reported structures form at given
experimental conditions.  The thermodynamic stability is determined by
the zero-Kelvin Gibbs free energy $G(\mu)=E-\mu N$, where $E$ is the
energy per Sr atom and $N$ is the number of Sr atoms.  $\mu$ is the
chemical potential of the Sr atom relative to our reference energy for
Sr. The extrinsic
quantities, such as energies $G$ and $E$, as well as the atom numbers
are measured per $(1\times1)$ unit cell of the silicon surface.  In
Fig.~\ref{fig:relenergyvscov} we show the adsorption energy $E$ per
$(1\times 1)$ surface unit cell.

The thermodynamically stable phases are determined by connecting the
points in Fig.~\ref{fig:relenergyvscov} by line segments and forming the
lower envelope. Each line segment corresponds to the coexistence of two
phases, denoted by $a$ and $b$, at the ends of the line segment with
energies $E_a$ and $E_b$ and $N_a$ and $N_b$ Sr atoms per $(1\times1)$
unit cell. For a coverage $N$ per $(1\times1)$ unit cell between $N_a$
and $N_b$, the energy is $E(N)=E_a+\frac{E_b-E_a}{N_b-N_a}(N-N_a)$. The
slope of the line segment, namely
$\mu=\frac{dE(N)}{dN}=\frac{E_a-E_b}{N_a-N_b}$, is the chemical
potential at which both phases $a$ and $b$ coexist. All structures not
contributing to the lower envelope are not thermodynamically stable at
low temperatures.  At higher temperatures they may be stabilized due to
entropic effects.

\begin{figure}
\includegraphics[angle=0,width=6cm,draft=false]{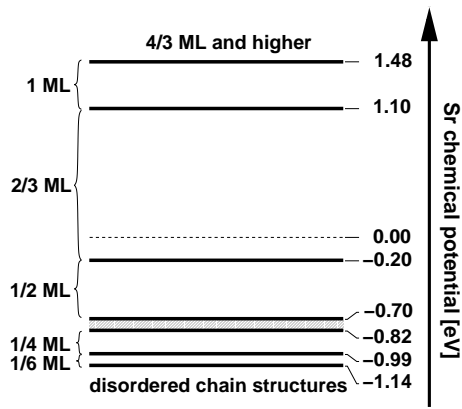}
\caption{One-dimensional phase diagram of Sr on Si(001) as a function of
the Sr chemical potential. The shaded region between the 1/4 and 1/2 ML
phase indicates disordered surface structures as described in
section~\ref{sec1.5}. Above the dashed line at a chemical potential of 0~eV
bulk Sr silicides are thermodynamically stable compared to surface
structures.}
\label{fig:chempot}
\end{figure}

The one-dimensional phase diagram is shown in Fig~\ref{fig:chempot}.
The region in between two lines correspond to the different surface
phases. The lines indicate the coexistence of the two
neighboring phases. 

We find thermodynamically stable phases at 1/6~ML and 1/4~ML.  We did
not fully explore the phase diagram below 1/6~ML. In this region we find
single chain structures as shown in Fig~\ref{fig:schematic1614}~(a) and
(b).  Entropic effects disorder the arrangement of chains at room
temperature.

At 1/4 ML we find an ordered structure of double chains. Beyond 1/4 ML
we predict nearly random arrangements of double vacancies.  The double
vacancy is stabilized relative to the single vacancy due to a
favorable dimer buckling.

The structure at 1/2 ML is a clear phase boundary for a wide range of
chemical potentials. 

The next phase boundary is found at a coverage of 2/3 ML with a
($3\times1$) reconstructed surface. Above 2/3 ML there are a number of
low-energy structures with various coverages and a $(n\times1)$
periodicity.   

Within the phase region of the 2/3~ML coverage, bulk Sr silicides become
thermodynamically stable as indicated by the dashed line in
Fig.~\ref{fig:chempot}. As mentioned above, we expect the onset of
silicide formation to be significantly delayed due to thin-film effects
as the bulk Sr silicides are highly incommensurate with the Si(001)
substrate.

It is of interest to compare our phase diagram with that obtained by
McKee et al.\cite{McKeeScience03} At around 600~C they determine three
line-compounds at 1/6~ML, 1/4~ML and 5/8~ML.  Our calculations reproduce
phase boundaries at 1/6~ML and 1/4~ML.  The next phase with a
$(3\times1)$ diffraction pattern is seen at $5/8=0.625$~ML,
which is close to the coverage of 2/3, where our calculations
predict a phase with the identical periodicity. The difference in
coverage corresponds to a change by one in 16 ad-atoms or 1/24~ML. This
difference may be attributed to occasional Sr vacancies, which help to
release strain. It seems surprising that no phase boundary is seen at
1/2~ML.  This may be due to the fact that on the one hand the
$(2\times1)$ structure develops continuously out of the multiple chains
structures below 1/2~ML and on the other hand it can be
transformed continuously into the $(3\times1)$ structure by introducing
thin stripes with local coverage of 1~ML.

The multiplicity of structures with low energy above a coverage of
2/3-ML suggests the presence of disordered structures at elevated
temperatures. McKee reports incommensurate structures beyond a
coverage of 5/8~ML.\cite{McKeeScience03}  

A similar multiplicity of structures is found for 1 ML. Our calculation
predicts a $(3\times1)$ structure as the most favorable. However we find
also $(4\times1)$, $(5\times1)$ and $(6\times1)$ reconstructions within
a window of 0.11~eV per Sr atom.

We did not extend our calculations beyond 4/3~ML, where an overlayer
of metallic Sr is formed. Therefore, our data does not necessarily
indicate the presence of a phase boundary at 4/3~ML.

\section{Conclusions}

In this paper we have investigated the surface structures of Sr adsorbed on
Si(001) as a function of coverage. We propose a theoretical phase
diagram by relating the phase boundaries at zero temperature to chemical
potentials, which can be converted into partial pressure and temperature
in thermal equilibrium. We predict phases at 1/6~ML, 1/4~ML, 1/2~ML,
2/3~ML and 1~ML.  Structural models are discussed for
all experimentally observed reconstructions except a $(3\times2)$
reconstructed layer attributed to a coverage of 1/3~ML.  The models are
explained in terms of structural templates and rationalized in terms of
their electronic structure. 

Our findings elucidate the chemistry of alkaline earth metals on Si(001)
and the phases of Sr on Si(001), which is expected to provide critical
information for the growth of one of the most promising high-K gate
oxide to date, namely SrTiO$_3$.

\begin{acknowledgments} 

We thank  A.~Dimoulas, J.~Fompeyrine, J.-P.~Loquet, R.A.~McKee and
G.~Norga for useful discussions.  This work has been funded by the
European Commission in the project "INVEST" (Integration of Very High-K
Dielectrics with CMOS Technology) and by the AURORA project of the
Austrian Science Fond.  Parts of the calculations have been performed on
the Computers of the ``Norddeutscher Verbund f\"ur Hoch- und
H\"ochstleistungsrechnen (HLRN)''.

\end{acknowledgments}
%

%

\begin{thebibliography}{99}
\bibitem{footnote1} The characteristic k-point spacing for bulk crystals
is calculated as $\sqrt[3]{\frac{(2\pi)^3}{\Omega\cdot N_k}}$, where
$\Omega$ is the volume of the unit cell and as
$\sqrt{\frac{(2\pi)^2}{A\cdot N_k}}$ for surfaces, where $A$ is the area
of the supercell. $N_k$ is the number of k-points per unit cell or
supercell. The actual number of k-points used in the calculation is
reduced relative to $N_k$ by exploiting time-inversion symmetry.
\bibitem{clemens} This energy is obtained from comparing the energies of
single, diagonal chains at 1/16~ML with and without buckling reversal.
Without introducing a buckling reversal, one positively charged Si atom
points towards the Sr atom. A buckling reversal  
allows to surround the Sr ad-atoms exclusively by negative Si atoms.  The
total energy of the cell with the reversed buckling has been corrected
by 0.06~eV as described above.
\bibitem{clemens1} The adsorption energy per $(1\times 1)$ unit cell is
defined as \(E=\big\{\frac{E[\mathrm{supercell}]-nE_0[\mathrm{m\
layer\mbox{-}Si\mbox{-}slab}]}{l}-E_0[\mathrm{Sr}] \big\}\cdot X\),
where $E[\mathrm{supercell}]$ is the total energy of the supercell used
for the specific surface reconstruction, $n$ is the number of $(1\times
1)$ surface unit cells contained in that supercell and $m$ is the
slab-thickness in units of silicon layers of the supercell. $l$
denotes the number of Sr atoms in the supercell and $X$ the Sr coverage
for that reconstruction.  This energy can be alternatively calculated
using the energy per Sr atom (see e.g.  Tab.~\ref{tab:ene}) multiplied
with the coverage $X$. The reference energies $E_0$ are listed in
Tab.~\ref{tab:reference}.
\end{thebibliography}
\end{document}